\documentclass[preprintnumbers,amsmath,amssymbm,prd]{revtex4}
\usepackage{epsfig}
\usepackage{graphicx}

\begin{document}
\title{A sufficient condition for the development of superradiant instabilities in charged black-hole spacetimes}
\author{Shahar Hod}
\affiliation{The Ruppin Academic Center, Emeq Hefer 40250, Israel}
\affiliation{ } \affiliation{The Hadassah Institute, Jerusalem
91010, Israel}
\date{\today}

\begin{abstract}
\ \ \ The physical and mathematical properties of charged black holes that are linearly coupled to 
charged massive scalar fields are studied analytically. 
In particular, we prove that, in the eikonal large-mass regime $M\mu\gg1$, the compact 
dimensionless inequality $\Phi_{\text{H}}>Q/M$ provides a sufficient condition 
for the development of superradiant instabilities in the curved black-hole spacetime [here 
$\{M,Q,\Phi_{\text{H}}\}$ are respectively the mass, the electric charge, and the horizon 
electrostatic potential of the central black hole and $\mu$ is the proper mass of the field]. 
The familiar charged Reissner-Nordstr\"om black hole does not satisfy this inequality. 
On the other hand, we explicitly prove that all charged Ay\'on-Beato-Garc\'ia (ABG) black-hole spacetimes satisfy 
this analytically derived sufficient condition and may therefore become superradiantly unstable to perturbations of 
charged massive scalar fields.  
\end{abstract}
\bigskip
\maketitle


\section{Introduction}

The physically intriguing superradiant amplification phenomenon implies that integer-spin (bosonic) 
fields that co-rotate with a spinning black hole of angular velocity $\Omega_{\text{H}}$ 
can extract its rotational energy if their proper frequencies lie in the superradiant interval 
\cite{Zel,PrTe,Viln,Noteunits}
\begin{equation}\label{Eq1}
0<\omega<m\Omega_{\text{H}}\  ,
\end{equation}
where the integer $m$ (with $m>0$) is the azimuthal harmonic index
of the bosonic field.

Using direct analytical techniques in the regime of linearized test fields \cite{Hod1,Hod2} 
and numerical computations in the non-linear regime of self-gravitating field configurations \cite{Her1,Her2,Gar1} 
it has been revealed that, thanks to the superradiant amplification 
phenomenon, spinning black-hole spacetimes with spatially regular horizons can support stationary co-rotating 
matter configurations which are made of minimally-coupled massive scalar (bosonic) fields whose 
frequencies, 
\begin{equation}\label{Eq2}
\omega=m\Omega_{\text{H}}\  ,
\end{equation}
are in resonance with the angular velocity that characterizes the horizon of the central 
supporting black hole. 

The physical significance of the composed black-hole-linearized-scalar-field 
configurations (`cloudy' black-hole-field bound-state configurations \cite{Hod1,Hod2,Her1,Her2,Gar1}), 
which are characterized by the resonance relation (\ref{Eq2}), 
stems from the fact that they mark the onset of superradiant instabilities of spinning 
black holes to perturbations of co-rotating massive scalar fields \cite{Hod1,Hod2,Her1,Her2,Gar1}. 

Charged bosonic fields that interact with a central charged black hole 
can also be superradiantly amplified if their proper frequencies lie in the superradiant interval \cite{Bekch}
\begin{equation}\label{Eq3}
0<\omega<q\Phi_{\text{H}}\  .
\end{equation}
Here $q$ is the charge coupling constant of the bosonic field and 
$\Phi_{\text{H}}$ is the electric potential at the outer horizon of the charged black hole \cite{Bekch} . 

Despite the fact that charged black holes can amplify charged bosonic fields in the frequency regime (\ref{Eq3}), 
it has been proved \cite{Hodnbt} that the charged Reissner-Nordstr\"om black hole cannot support stationary matter 
configurations which are made of minimally coupled charged massive 
scalar fields \cite{Notecbb,Dego,Herrec,Hodexp,Lim} [this property of charged Reissner-Nordstr\"om 
black holes should be contrasted with the fact that spinning Kerr black holes can 
support linearized scalar field configurations (massive scalar `clouds') \cite{Hod1,Hod2,Her1,Her2,Gar1} which 
are characterized by the frequency relation (\ref{Eq2})]. 
In accord with this finding, charged Reissner-Nordstr\"om black holes are known to be 
stable to perturbations of minimally coupled charged massive scalar fields \cite{Hodnbt}. 

On the other hand, it has recently been demonstrated in the physically interesting works \cite{PL,PL2} (see 
also \cite{Hodabg}) that charged Ay\'on-Beato-Garc\'ia (ABG) black holes \cite{ABG} 
may become superradiantly unstable when coupled to charged massive scalar fields in the frequency regime  (\ref{Eq3}).  

The main goal of the present paper is to explore the 
physical and mathematical properties of composed charged-black-hole-charged-massive-scalar-field systems. 
In particular, motivated by the intriguing fact that some charged black-hole spacetimes are stable to perturbations of 
charged massive scalar fields \cite{Hodnbt}, whereas other charged black-hole spacetimes 
may develop superradiant instabilities \cite{ABG,PL,PL2,Hodabg}, 
we here raise the following physically important question: 
Is it possible to derive a sufficient condition for the development of superradiant instabilities in curved spacetimes  
of charged black holes?

Interestingly, we shall explicitly prove below that, in the eikonal large-mass (or equivalently, large-frequency) 
regime $M\mu\gg1$, the compact dimensionless inequality 
\begin{equation}\label{Eq4}
\Phi_{\text{H}}>{{|Q|}\over{M}}\
\end{equation}
provides a sufficient condition for the existence of composed 
charged-black-hole-linearized-charged-massive-scalar-field 
bound-state configurations. 
Here $\{M,Q\}$ are respectively the mass and the electric charge of the central black hole and 
$\mu$ is the proper mass of the supported scalar field. 

The stationary charged scalar clouds, which are 
characterized by the critical (marginal) resonant frequency $\omega=q\Phi_{\text{H}}$ 
for the superradiant amplification phenomenon in the charged black-hole spacetime [see Eq. (\ref{Eq3})], 
mark the onset of superradiant instabilities in the supporting black-hole spacetime. 
Thus, in the dimensionless large-mass $M\mu\gg1$ regime, 
the inequality (\ref{Eq4}), to be derived below, 
provides a sufficient condition for the development of superradiant instabilities of charged massive scalar fields 
in charged black-hole spacetimes. 

\section{Description of the system}

We consider a physical system which is composed of a minimally coupled linearized scalar field 
of proper mass $\mu$ and electric charge $q$ \cite{Notemuq} that interacts with a central black hole 
of mass $M$ and electric charge $Q$ \cite{Noteqp}. 
The curved black-hole spacetime is characterized by the spherically symmetric 
line element \cite{Chan,Notesch,Notefg,Dym}
\begin{equation}\label{Eq5}
ds^2=-f(r)dt^2+{1\over{f(r)}}dr^2+r^2(d\theta^2+\sin^2\theta d\varphi^2)\  .
\end{equation}
The horizon radii of the black hole are determined by the roots of the dimensionless 
metric function $f(r)$:
\begin{equation}\label{Eq6}
f(r=r_{\text{H}})=0\  . 
\end{equation}
An asymptotically flat spacetime is characterized by the functional behavior
\begin{equation}\label{Eq7}
f(r\to\infty)\to1\  . 
\end{equation}

The Klein-Gordon wave equation \cite{HodPirpam,Stro,HodCQG2}
\begin{equation}\label{Eq8}
[(\nabla^\nu-iqA^\nu)(\nabla_{\nu}-iqA_{\nu}) -\mu^2]\Psi=0\
\end{equation}
governs the the dynamics of the charged massive scalar field $\Psi$ in the curved spacetime (\ref{Eq5}). 
Here $A_{\nu}=-\delta_{\nu}^{0}\Phi(r)$ is the electromagnetic 
potential of the charged black hole. 
Using the tortoise radial coordinate $y=y(r)$, which is defined by the relation \cite{Notery} 
\begin{equation}\label{Eq9}
dy={{dr}\over{f(r)}}\  ,
\end{equation}
one can write Eq. (\ref{Eq8}) in the mathematically compact form
\begin{equation}\label{Eq10}
{{d^2R_{lm}}\over{dy^2}}-VR_{lm}=0\  ,
\end{equation}
where we have used here the scalar field decomposition \cite{Notelm}
\begin{equation}\label{Eq11}
\Psi(t,r,\theta,\varphi)=\int\sum_{lm} {{1}\over{r}}R_{lm}(r)
Y_{lm}(\theta,\varphi)e^{im\varphi}e^{-i\omega t}d\omega\  .
\end{equation}
The angular scalar eigenfunctions $Y_{lm}(\theta,\varphi)$ in (\ref{Eq11}) 
are the spherical harmonic functions which are 
characterized by the familiar discrete angular eigenvalues $\lambda_l=l(l+1)$ \cite{Noteom}. 
The composed black-hole-field radial potential in 
the Schr\"odinger-like ordinary differential equation (\ref{Eq10}) is given by the 
functional expression \cite{HodPirpam,Stro,HodCQG2}
\begin{eqnarray}\label{Eq12}
V[r(y)]=f(r)\Big[\mu^2+{{1}\over{r}}{{df(r)}\over{dr}}+{{l(l+1)}\over{r^2}}\Big]-
\big[\omega-q\Phi(r)\big]^2\  .
\end{eqnarray}

The composed charged-black-hole-charged-massive-scalar-field cloudy configurations, 
which mark the onset of superradiant instabilities in the charged black-hole spacetime (\ref{Eq5}), 
are characterized by the critical (marginal) resonant frequency \cite{Notemss}
\begin{equation}\label{Eq13}
\omega=q\Phi_{\text{H}}\ 
\end{equation}
for the superradiant amplification phenomenon of charged bosonic fields in a charged 
black-hole spacetime [see Eq. (\ref{Eq3})]. 

The Schr\"odinger-like differential equation (\ref{Eq10}) is supplemented by the asymptotic 
large-$r$ (large-$y$) boundary condition \cite{Hodnbt}
\begin{equation}\label{Eq14}
R \sim e^{-\sqrt{\mu^2-\omega^2}y}\ \ \ \ \text{for}\ \ \ \ r\to\infty\ \ (y\rightarrow \infty)\
\end{equation}
which, for field frequencies in the regime 
\begin{equation}\label{Eq15}
\omega^2<\mu^2\  ,
\end{equation}
characterizes spatially bounded (normalizable) scalar configurations. 
In addition, the physically motivated boundary condition \cite{Noterh} 
\begin{equation}\label{Eq16}
R \sim e^{-i (\omega-q\Phi_{\text{H}})y}\ \ \ \ \text{for}\ \ \ \ r\to r_{\text{H}}\ \ (y\rightarrow -\infty)\
\end{equation}
characterizes scalar waves that are purely ingoing (as measured by a comoving observer) 
at the outer horizon of the charged black hole.

It is worth emphasizing that, as explicitly proved in \cite{Hodnbt}, 
not all charged black holes can support spatially regular charged massive scalar clouds. 
In the next section we shall use analytical techniques, which are valid in the dimensionless 
large-mass $M\mu\gg1$ regime, in order to derive a physically useful sufficient condition for the existence of 
composed charged-black-hole-linearized-charged-massive-scalar-field bound-state configurations in curved spacetimes. 

\section{A sufficient condition for the development of superradiant instabilities in the composed 
charged-black-hole-charged-massive-scalar-field system}

In the present section we shall derive, using analytical techniques, a mathematically compact inequality which, 
in the eikonal large-mass $M\mu\gg1$ regime, provides a sufficient condition 
for the development of superradiant instabilities in the charged black-hole spacetime (\ref{Eq5}). 

To this end, we first note that the radial potential (\ref{Eq12}) of a stationary charged massive scalar field 
with the critical (marginal) resonant frequency $\omega=q\Phi_{\text{H}}$ 
is characterized by the near-horizon relation [see Eqs. (\ref{Eq6}), (\ref{Eq12}), and (\ref{Eq13})]
\begin{equation}\label{Eq17}
V(r\to r^+_{\text{H}})\to0\
\end{equation}
and by the asymptotic large-$r$ relation [see Eqs. (\ref{Eq7}), (\ref{Eq12}), (\ref{Eq13}), and (\ref{Eq15})]
\begin{equation}\label{Eq18}
V(r\to\infty)\to \mu^2-(q\Phi_{\text{H}})^2>0\  .
\end{equation}
 
As discussed above, the onset of superradiant instabilities in charged black-hole spacetimes is marked by the presence 
of marginally-stable (stationary) bound-state matter configurations, which are made of 
charged massive scalar fields with the critical resonant frequency (\ref{Eq13}), 
that are supported by the central black hole. A necessary condition for the existence of these stationary 
bound-state field configurations is provided by the requirement that the 
black-hole-field effective radial potential (\ref{Eq12}) has a binding well. 
In particular, taking cognizance of the asymptotic behaviors (\ref{Eq17}) and (\ref{Eq18}) of the composed 
black-hole-field radial potential (\ref{Eq12}), one deduces that the inequality
\begin{equation}\label{Eq19}
\text{min}_r\{V(r)\}<0\
\end{equation}
provides a necessary condition for the existence of the 
marginally-stable linearized scalar clouds in the 
charged black-hole spacetime (\ref{Eq5}) \cite{Notesimr,ChunHer,Hodca,Herkn,Hodjp,Hodar}. 

We shall now prove that a sufficient condition for the existence of stationary charged scalar clouds in charged 
black-hole spacetimes can be derived analytically in the dimensionless large-mass regime 
[or equivalently, in the asymptotic large-frequency $\omega r_{\text{H}}\gg1$ regime]
\begin{equation}\label{Eq20}
\mu r_{\text{H}}\gg1\  ,
\end{equation}
in which case the composed black-hole-charged-massive-scalar-field radial potential (\ref{Eq12})
can be written in the compact mathematical form
\begin{eqnarray}\label{Eq21}
V(r)=\Big\{f(r)\cdot\mu^2-q^2\big[\Phi_{\text{H}}-\Phi(r)\big]^2\Big\}\cdot
\{1+O[(\mu r_{\text{H}})^{-2}]\}\ \ \ \ \text{for}\ \ \ \ \mu r_{\text{H}}\gg1\  .
\end{eqnarray}

In the eikonal regime (\ref{Eq20}) of large black-hole-field masses 
the Schr\"odinger-like ordinary differential equation (\ref{Eq10}) of the charged massive scalar field 
is characterized by the familiar second-order WKB quantization condition \cite{WKB1,WKB2,WKB3}
\begin{equation}\label{Eq22}
\int_{y_{t_-}}^{y_{t_+}}dy\sqrt{-V(y)}=\big(n+{1\over2}\big)\cdot\pi\ 
\ \ \ ; \ \ \ \ n=0,1,2,...\  ,
\end{equation}
where the radial integration boundaries, which are determined by the relations
\begin{equation}\label{Eq23}
V(y_{t_-})=V(y_{t_+})=0\  ,
\end{equation}
are the classical turning points of the composed charged-black-hole-charged-massive-scalar-field 
binding potential (\ref{Eq21}). 
The WKB resonance equation (\ref{Eq22}) can be expressed in the form [see Eq. (\ref{Eq9})]
\begin{equation}\label{Eq24}
\int_{r_{t_-}}^{r_{t_+}}dr {{\sqrt{-V(r)}}\over{f(r)}}=\big(n+{1\over2}\big)\cdot\pi\ 
\ \ \ ; \ \ \ \ n=0,1,2,...\  .
\end{equation}

Using the radial functional expansions 
\begin{equation}\label{Eq25}
V(r)=V(r_{\text{min}})+{1\over2}\Big({{d^2V}\over{dr^2}}\Big)_{r=r_{\text{min}}}\cdot(r-r_{\text{min}})^2\
\end{equation}
and
\begin{equation}\label{Eq26}
f(r)=f(r_{\text{min}})+\Big({{df}\over{dr}}\Big)_{r=r_{\text{min}}}\cdot(r-r_{\text{min}})\
\end{equation}
for the black-hole-field binding potential and the metric function in the vicinity of the 
radial minimum point of the potential, for which 
\begin{equation}\label{Eq27}
\Big({{dV}\over{dr}}\Big)_{r=r_{\text{min}}}=0\  ,
\end{equation} 
and defining the dimensionless radial coordinate
\begin{equation}\label{Eq28}
z=\sqrt{-{{\big({{d^2V}\over{dr^2}}\big)_{r=r_{\text{min}}}}\over{2V(r_{\text{min}})}}}\cdot(r-r_{\text{min}})\  ,
\end{equation}
one can express the WKB resonance condition (\ref{Eq24}) in the form
\begin{equation}\label{Eq29}
-{{\sqrt{2}V(r_{\text{min}})}\over{f(r_{\text{min}})\sqrt{{{\big({{d^2V}\over{dr^2}}\big)_{r=r_{\text{min}}}}}}}}
\int_{-1}^{1}dz\sqrt{1-z^2}
=\big(n+{1\over2}\big)\cdot\pi\ \ \ \ ; \ \ \ \ n=0,1,2,...\  .
\end{equation}
From Eq. (\ref{Eq29}) one finds the compact mathematical relation \cite{Noteintegral,Notent}
\begin{equation}\label{Eq30}
V(r_{\text{min}})=-\sqrt{2}\cdot\big(n+{1\over2}\big)\cdot f(r_{\text{min}})
\cdot\sqrt{{{\Big({{d^2V}\over{dr^2}}\Big)_{r=r_{\text{min}}}}}}<0\  .
\end{equation}
Taking cognizance of Eqs. (\ref{Eq21}) and (\ref{Eq30}), one deduces the characteristic 
dimensionless ratio
\begin{equation}\label{Eq31}
-{{V(r_{\text{min}})}\over{\mu^2}}=O\big[(\mu r_{\text{H}})^{-1}\big]\ll1
\ \ \ \ \text{for}\ \ \ \ \mu r_{\text{H}}\gg1\
\end{equation}
for the effective binding potential of the composed black-hole-massive-field cloudy configurations 
in the eikonal large-mass regime (\ref{Eq20}) \cite{Notewit}. 

Thus, in the eikonal large-mass regime (\ref{Eq20}), the two coupled functional relations (\ref{Eq27}) and (\ref{Eq31}) 
provide a sufficient condition for the existence of 
marginally-stable (stationary) linearized scalar clouds in the charged black-hole spacetime (\ref{Eq5}). 
These composed black-hole-linearized-scalar-field cloudy configurations mark the boundary between stable and 
superradiantly unstable spacetimes. 

From Eqs. (\ref{Eq21}), (\ref{Eq27}), and (\ref{Eq31}) one finds that, 
in the dimensionless large-mass regime (\ref{Eq20}), composed 
charged-black-hole-charged-massive-scalar-field cloudy configurations, which mark the onset of 
the superradiant instability phenomenon in the charged black-hole spacetime (\ref{Eq5}), are 
characterized by the functional relation
\begin{equation}\label{Eq32}
1+{{\big({{df}\over{dr}}\big)_{r=r_{\text{min}}}\cdot
\big[\Phi_{\text{H}}-\Phi(r_{\text{min}})\big]}\over{2f(r_{\text{min}})
\cdot\big({{d\Phi}\over{dr}}\big)_{r=r_{\text{min}}}}}
=-{{C}\over{\mu r_{\text{H}}}}\to0^-
\ \ \ \ \text{for}\ \ \ \ \mu r_{\text{H}}\gg1\  ,
\end{equation}
where $0<C=O(1)$. 

We shall now derive a sufficient condition for the validity of the large-mass relation (\ref{Eq32}). 
To this end, we shall analyze the asymptotic functional behaviors of the radially-dependent dimensionless function 
\begin{equation}\label{Eq33}
{\cal F}(r)\equiv 1+{{{{df}\over{dr}}\cdot\big[\Phi_{\text{H}}-\Phi(r)\big]}
\over{2f(r)\cdot{{d\Phi}\over{dr}}}}\
\end{equation}
in the two opposite limits $r/r_{\text{H}}\to1^+$ and $r\to\infty$. 


In the near-horizon $r=r_{\text{H}}+\Delta r$ region with $0<\Delta r/r_{\text{H}}\ll1$, 
one can substitute into (\ref{Eq33}) the functional expansion
\begin{equation}\label{Eq34}
\Phi(r)=\Phi_{\text{H}}+\Big({{d\Phi}\over{dr}}\Big)_{r=r_{\text{H}}}\cdot\Delta r\
\end{equation}
for the electrostatic potential and
\begin{equation}\label{Eq35}
f(r)=\Big({{df}\over{dr}}\Big)_{r=r_{\text{H}}}\cdot\Delta r\
\end{equation}
for the metric function of non-extremal black holes to find the simple near-horizon behavior
\begin{equation}\label{Eq36}
{\cal F}(r=r_{\text{H}}+\Delta r)={1\over2}\cdot[1+O(\Delta r/r_{\text{H}})]>0\  ,
\end{equation}
which is a positive definite quantity. 

On the other hand, substituting into (\ref{Eq33}) the radial relations  
\begin{equation}\label{Eq37}
f(r\to\infty)\to 1-{{2M}\over{r}}\
\end{equation}
and
\begin{equation}\label{Eq38}
\Phi(r\to\infty)\to {{Q}\over{r}}\ 
\end{equation}
for the weak-field asymptotic functional behaviors of the metric function and the electrostatic potential [here $\{M,Q\}$ 
are respectively the asymptotically measured mass and the electric charge of the 
black-hole spacetime], one finds the large-$r$ functional behavior 
\begin{equation}\label{Eq39}
{\cal F}(r\to\infty)\to1-{{M\Phi_{\text{H}}}\over{Q}}\ 
\end{equation}
of the dimensionless function (\ref{Eq33}). 
The asymptotic large-$r$ expression (\ref{Eq39}) 
is negative,
\begin{equation}\label{Eq40}
{\cal F}(r\to\infty)<0\ \ \ \ \text{for}\ \ \ \ \ \Phi_{\text{H}}>{{Q}\over{M}}\  ,
\end{equation}
for charged black holes which are characterized by the inequality 
$M\Phi_{\text{H}}/Q>1$.

Interestingly, and most importantly for our analysis, from Eqs. (\ref{Eq36}) and (\ref{Eq40}) one deduces 
that, for black holes that respect the inequality 
\begin{equation}\label{Eq41}
\Phi_{\text{H}}>{{Q}\over{M}}\cdot\{1+O[(\mu r_{\text{H}})^{-1}]\}\ \ \ \ \text{for}\ \ \ \ \mu r_{\text{H}}\gg1\  ,
\end{equation}
the analytically derived WKB equation (\ref{Eq32}) must have at least one radial solution 
which is located between the black-hole horizon and spatial infinity. 

Thus, charged black holes that respect the dimensionless inequality (\ref{Eq41}) can support 
stationary charged massive scalar clouds that are characterized by the resonant frequency (\ref{Eq13}). 
The existence of these composed charged-black-hole-linearized-charged-massive-scalar-field bound-state 
configurations marks the onset of superradiant instabilities in the curved black-hole spacetime. 

It is physically interesting to note that the canonical family of 
charged Reissner-Nordstr\"om black-hole spacetimes, which are known to be stable to perturbations 
of charged massive scalar fields \cite{Hodnbt}, does not satisfy the inequality (\ref{Eq41}). 
In the next section we shall explicitly demonstrate that other families of charged black-hole spacetimes 
may respect the analytically derived dimensionless inequality (\ref{Eq41}) and may therefore become superradiantly 
unstable to perturbations of charged massive scalar fields in the dimensionless large-mass regime (\ref{Eq20}).  

\section{Superradiant instabilities of charged Ay\'on-Beato-Garc\'ia (ABG) black holes}

In the present section we shall use our analytically derived sufficient condition (\ref{Eq41}) 
in order to prove that all charged Ay\'on-Beato-Garc\'ia (ABG) black holes \cite{ABG,PL,PL2,Hodabg} 
may become superradiantly unstable when coupled to charged massive scalar fields 
in the eikonal large-mass regime (\ref{Eq20}). 

The ABG black-hole spacetime, which is characterized by the curved line element (\ref{Eq5}) 
with the radially-dependent metric function \cite{ABG,PL}
\begin{equation}\label{Eq42}
f^{\text{ABG}}(r)=1-{{2Mr^2}\over{(r^2+Q^2)^{3/2}}}+{{Q^2r^2}\over{(r^2+Q^2)^2}}\  ,
\end{equation} 
describes a spatially regular solution of the coupled Einstein-non-linear-electrodynamics field 
equations (see \cite{ABG,PL} and references therein for details). 
The electric potential of the ABG spacetime (\ref{Eq42}) is given by the non-trivial dimensionless 
functional expression \cite{ABG,PL}
\begin{equation}\label{Eq43}
\Phi^{\text{ABG}}(r)={{r^5}\over{2Q}}\Big[{{3M}\over{r^5}}+{{2Q^2}\over{(r^2+Q^2)^3}}-{{3M}
\over{(r^2+Q^2)^{5/2}}}\Big]\  .
\end{equation}

From Eqs. (\ref{Eq6}) and (\ref{Eq42}) one obtains the ratio
\begin{equation}\label{Eq44}
{\bar M}\equiv{{M}\over{r_{\text{H}}}}={{1+3{\bar Q}^2+{\bar Q}^4}\over
{2\sqrt{1+{\bar Q}^2}}}\  ,
\end{equation}
where 
\begin{equation}\label{Eq45}
{\bar Q}\equiv{{Q}\over{r_{\text{H}}}}\ 
\end{equation}
is the dimensionless charge parameter of the black hole. 

Substituting Eqs. (\ref{Eq44}) and (\ref{Eq45}) into Eq. (\ref{Eq43}) one obtains the functional expression 
\begin{equation}\label{Eq46}
\Phi_{\text{H}}={{3(1+3{\bar Q}^2+{\bar Q}^4)}
\over{4{\bar Q}\sqrt{1+{\bar Q}^2}}}-
{{3+5{\bar Q}^2+3{\bar Q}^4}\over{4{\bar Q}(1+{\bar Q}^2)^3}}\
\end{equation}
for the electric potential at the horizon of the charged ABG black hole. 

A direct inspection of the dimensionless function [see Eqs. (\ref{Eq41}), (\ref{Eq44}), (\ref{Eq45}), and (\ref{Eq46})]
\begin{equation}\label{Eq47}
{\cal G}({\bar Q})\equiv {{M\Phi_{\text{H}}\over{Q}}}\
\end{equation}
reveals that it increases monotonically with the value of the black-hole charge parameter $|{\bar Q}|$. 
In particular, one finds the characteristic inequality
\begin{equation}\label{Eq48}
{\cal G}({\bar Q})>1
\end{equation}
for all values of the dimensionless charge parameter ${\bar Q}$ with the minimum relation   
\begin{equation}\label{Eq49}
\text{min}\{{\cal G}({\bar Q})\}={{23}\over{16}}\ \ \ \ \text{for}\ \ \ \ {\bar Q}\to0\  .
\end{equation}

Taking cognizance of Eqs. (\ref{Eq47}), (\ref{Eq48}), and (\ref{Eq49}), 
one deduces that all charged ABG 
black holes satisfy the analytically derived sufficient condition (\ref{Eq41}) and may therefore 
become superradiantly unstable to perturbations of charged massive scalar fields 
in the eikonal large-mass regime (\ref{Eq20}). 
It is physically interesting to note that this analytically deduced property of the charged ABG black-hole spacetimes 
is in accord with the findings of recent numerical \cite{PL,PL2} and analytical \cite{Hodabg} studies that explored 
the physical properties of the composed ABG-black-hole-charged-massive-scalar-field system. 

\section{Summary}

It is well established that, as a consequence of 
the physically interesting superradiant amplification phenomenon of bosonic fields in rotating spacetimes \cite{Zel,PrTe,Viln}, 
Kerr black holes can support scalar clouds \cite{Hod1,Hod2,Her1,Her2,Gar1}, linearized co-rotating massive scalar fields that mark the onset of 
superradiant instabilities in the corresponding spinning spacetime. 

On the other hand, it has been explicitly proved \cite{Hodnbt} that, despite the fact that 
charged bosonic fields can be superradiantly amplified by extracting electromagnetic energy from 
charged black holes \cite{Bekch}, charged Reissner-Nordstr\"om black holes cannot support linearized 
charged scalar clouds. As a direct consequence, Reissner-Nordstr\"om black-hole spacetimes are known 
to be stable to perturbations of minimally coupled charged massive scalar fields \cite{Hodnbt}. 

Motivated by the intriguing fact that some charged black-hole spacetimes are stable to perturbations of 
charged massive scalar fields \cite{Hodnbt}, whereas other charged black-hole spacetimes 
may develop superradiant instabilities \cite{ABG,PL,PL2,Hodabg}, in the present compact paper we have raised 
the following physically important question: 
Is it possible to derive a generic sufficient condition that guarantees the existence of superradiant instabilities 
in curved spacetimes of charged black holes?

In order to address this interesting question we have studied, using analytical techniques, 
the physical and mathematical properties of composed charged-black-hole-linearized-charged-massive-scalar-field systems. 
In particular, we have proved that, in the eikonal large-mass $\mu r_{\text{H}}\gg1$ regime, 
the dimensionless inequality [see Eqs. (\ref{Eq20}) and (\ref{Eq41})]
\begin{equation}\label{Eq50}
\Phi_{\text{H}}>{{Q}\over{M}}\ \ \ \ \text{for}\ \ \ \ \mu r_{\text{H}}\gg1\
\end{equation}
provides a sufficient condition for the existence of stationary bound-state charged scalar clouds that 
are characterized by the critical (marginal) resonant frequency $\omega=q\Phi_{\text{H}}$ 
for the superradiant amplification phenomenon in the charged black-hole spacetime (\ref{Eq5}). 

Interestingly, taking cognizance of the fact that marginally-stable (stationary) scalar clouds 
mark the onset of superradiant instabilities in curved black-hole spacetimes \cite{Hod1,Hod2,Her1,Her2,Gar1}, one concludes that, in the dimensionless large-mass $M\mu\gg1$ regime, 
the physical property (\ref{Eq50}) 
provides a sufficient condition for the development of superradiant instabilities 
in charged black-hole spacetimes. 


\bigskip
\noindent
{\bf ACKNOWLEDGMENTS}
\bigskip

This research is supported by the Carmel Science Foundation. I thank
Yael Oren, Arbel M. Ongo, Ayelet B. Lata, and Alona B. Tea for
stimulating discussions.



\begin{thebibliography}{99}
\bibitem{Noteunits} We use natural units in which $G=c=\hbar=1$.
\bibitem{Notecbb} It is worth noting that a charged Reissner-Nordstr\"om black hole 
may become superradiantly unstable to perturbations of charged bosonic fields 
if it is surrounded by a mirror (a reflecting surface) that prevents the amplified 
bosonic field, which is characterized by the superradiant frequency relation (\ref{Eq3}), 
from radiating its energy to infinity (see \cite{Dego,Herrec,Hodexp,Lim} for details).
\bibitem{Notemuq} Note that the parameters $\mu$ and $q$ stand 
respectively for $\mu/\hbar$ and $q/\hbar$. Hence, these field parameters 
have the dimensions of length$^{-1}$.
\bibitem{Noteqp} One may assume, without loss of generality, the relations $Q>0$ and $q>0$ for the 
electric charges of the black hole and the massive scalar field.
\bibitem{Notesch} Here $\{t,r,\theta,\varphi\}$ are the familiar Schwarzschild coordinates.
\bibitem{Notefg} Note that if the electromagnetic part of the Lagrangian ${\cal L}={\cal L}(F)$ 
is a function of the electromagnetic invariant $F=F_{\mu\nu}F^{\mu\nu}$ then the energy momentum 
tensor of the physical system is characterized by the functional relation $T^{t}_{t}=T^{r}_{r}$ 
[or equivalently $p_r=-\rho$, where $\{\rho,p_r\}$ are respectively the energy density and the radial pressure] 
\cite{Dym}, in which case the non-linearly coupled Einstein-matter field equations imply that 
the spherically symmetric black-hole spacetime is characterized by the curved line element (\ref{Eq5}) 
with $g_{rr}=(g_{tt})^{-1}$ \cite{Dym}.
\bibitem{Notery} Note that the radial coordinate $r\in[r_{\text{H}},\infty]$ is mapped by 
the differential relation (\ref{Eq9}) to the radial coordinate $y\in[-\infty,\infty]$.
\bibitem{Notelm} Note that the integers $l$ and $m$ are respectively 
the spherical harmonic index and the azimuthal harmonic index (with $l\geq|m|$) of the field mode.
\bibitem{Noteom} For brevity, we shall henceforth omit the harmonic indices $l$ and $m$ 
of the scalar field mode.
\bibitem{Notemss} Note that stationary field modes, which are characterized by the critical 
frequency $\omega=q\Phi_{\text{H}}$ [see Eq. (\ref{Eq13})] 
with $\Im\omega=0$, mark the boundary between stable (exponentially decaying) and
unstable (exponentially growing) resonant modes of the massive bosonic fields \cite{Hod1,Hod2,Her1,Her2,Gar1}.
\bibitem{Noterh} We shall henceforth use the notation $r_{\text{H}}$ for the radius of the 
outer horizon of the central charged black hole.
\bibitem{Notesimr} See \cite{ChunHer,Hodca,Herkn,Hodjp,Hodar} and references therein 
for a physically interesting situation in which non-minimally coupled static 
scalar fields are supported by central black holes with spatially regular horizons.
\bibitem{Noteintegral} Here we have used the simple relation $\int_{-1}^{1}dz \sqrt{1-z^2}={{\pi}/{2}}$.
\bibitem{Notent} Note that $f(r_{\text{min}})\cdot\sqrt{(d^2V/dr^2)_{r=r_{\text{min}}}}=
\sqrt{(d^2V/dy^2)_{y=y_{\text{min}}}}$.
\bibitem{Notewit} From Eqs. (\ref{Eq21}), (\ref{Eq25}), and (\ref{Eq31}) one finds that, 
in the dimensionless large-mass regime $M\mu\gg1$, 
the effective radial widths $\Delta r\equiv r_{t_+}-r_{t_-}$ of the stationary massive scalar clouds 
scale as $\Delta r/r_{\text{H}}\propto 1/\sqrt{M\mu}$, where $\{r_{t_+},r_{t_-}\}$ are the 
radial zeros (radial turning points) of the black-hole-field effective radial potential. 
One therefore finds that the stationary charged scalar clouds, which are supported in the charged 
black-hole spacetime (\ref{Eq5}), can be made arbitrarily thin in the dimensionless 
large-mass $M\mu\gg1$ regime. 
In particular, in the eikonal large-mass $M\mu\gg1$ regime, the massive scalar clouds are 
sharply centered around $r=r_{\text{min}}$ as defined by Eq. (\ref{Eq27}).
\end{thebibliography}
\end{document}